\documentclass[%
reprint,
superscriptaddress,
showpacs,
amsmath,amssymb,
aps,
]{revtex4-2}
\usepackage{graphicx}
\usepackage{dcolumn}
\usepackage{bm}
\usepackage{CJKutf8}
\usepackage{color}
\usepackage{amssymb}
\usepackage{amsfonts}
\usepackage{palatino}
\usepackage[colorlinks,urlcolor=blue,linkcolor=blue,citecolor=blue,anchorcolor=blue]{hyperref}
\makeatletter

\newcommand{\Rmnum}[1]{\expandafter\@slowromancap\romannumeral #1@}

\makeatother
\begin{document}

\begin{CJK*}{UTF8}{gbsn}

\preprint{APS/123-QED}

\title{$\pi$ mode lasing in the non-Hermitian Floquet topological system}

\author{Shuang Shen}
\affiliation{Key Laboratory for Physical Electronics and Devices, Ministry of Education, School of Electronic Science and Engineering, Xi'an Jiaotong University, Xi'an 710049, China}

\author{Yaroslav V. Kartashov}
\affiliation{Institute of Spectroscopy, Russian Academy of Sciences, Troitsk, Moscow, 108840,	Russia}

\author{Yongdong Li}
\affiliation{Key Laboratory for Physical Electronics and Devices, Ministry of Education, School of Electronic Science and Engineering, Xi'an Jiaotong University, Xi'an 710049, China}

\author{Meng Cao}
\affiliation{Key Laboratory for Physical Electronics and Devices, Ministry of Education, School of Electronic Science and Engineering, Xi'an Jiaotong University, Xi'an 710049, China}

\author{Yiqi Zhang}
\email{zhangyiqi@xjtu.edu.cn}
\affiliation{Key Laboratory for Physical Electronics and Devices, Ministry of Education, School of Electronic Science and Engineering, Xi'an Jiaotong University, Xi'an 710049, China}

\date{\today}

\begin{abstract}
\noindent
$\pi$ modes are unique topological edge states appearing in Floquet systems with periodic modulations of the underlying lattice structure in evolution variable, such as dynamically modulated Su-Schrieffer-Heeger (SSH) lattices. These edge states are anomalous states usually appearing between Floquet replicas of the same band, even if standard topological index remains zero for this band. While linear and nonlinear $\pi$ modes were observed in conservative systems, they have never been studied in nonlinear regime in the non-Hermitian systems with structured gain and losses. Here we show that SSH waveguide array with periodically oscillating waveguide positions in propagation direction and with parity-time symmetric refractive index landscape, can support $\pi$ modes that are damped or amplified at different ends of the array. By including nonlinearity and nonlinear absorption into our continuous system, we achieve stable lasing in $\pi$ mode at one end of the array. The representative feature of this system is that lasing in it is thresholdless and it occurs even at low gain-loss amplitudes. The degree of localization of lasing $\pi$ modes can be flexibly controlled by the amplitude of transverse waveguide oscillations. This work therefore introduces a new type of topological Floquet laser and a route to manipulation of $\pi$ modes by structured gain and losses.
\end{abstract}

\maketitle

\end{CJK*}

\section{Introduction}

The extension of the concept of topological insulators, originally discovered in solid-state physics~\cite{hasan.rmp.82.3045.2010, qi.rmp.83.1057.2011}, to the realm of photonic systems~\cite{lu.np.8.821.2014, ozawa.rmp.91.015006.2019, kim.lsa.9.130.2020, segev.nano.10.425.2021, ma.pi.1.R02.2022, price.jp.4.032501.2022, zhang.nature.618.687.2023} has allowed to predict and observe experimentally many previously elusive topological phases, introduce new mechanisms of formation of topological edge states, and has revealed a number of new possible practical applications of topological systems, particularly at optical frequencies~\cite{rechtsman.nature.496.196.2013}. Different branches of topological photonics that study the impact of nonlinearity on topological edge states~\cite{smirnova.apr.7.021306.2020}, quantum effects in topological systems \cite{yan.aom.2001739.2021}, and the impact of non-Hermitian effects~\cite{ota.nano.9.547.2020, parto.nano.10.403.2021, wang.jo.23.123001.2021, ding.nrp.4.745.2022, yan.nano.12.2247.2023, nasari.ome.13.870.2023} demonstrate rapid development. Particularly intriguing mechanism of formation of topological edge states has been encountered in systems periodically modulated in the evolution variable (for example, in the direction of light propagation). Such Floquet systems~\cite{rudner.prx.3.031005.2013, rudner.nrp.2.229.2020} allow to observe novel anomalous topological phases, where edge states of topological origin emerge between Floquet replicas of the same band, appearing exclusively due to longitudinal refractive index modulation~\cite{maczewsky.nc.8.13756.2017, mukherjee.nc.8.13918.2017}. When periodic longitudinal modulation is applied to the simplest Su-Schrieffer-Heeger (SSH) model of the one-dimensional topological system~\cite{asboth.book.2016, malkova.ol.34.1633.2009}, it also can give rise to novel anomalous topological edge states called $\pi$ modes, that are qualitatively different from conventional topological ``zero-energy" states~\cite{dallago.pra.92.023624.2015, fruchart.prb.93.115429.2016, zhang.acs.4.2250.2017, petracek.pra.101.033805.2020, cheng.prl.122.173901.2019, song.lpr.15.2000584.2021, sidorenko.prr.4.033184.2022}. 

Inclusion of nonlinearity in topological system adds desired tunability, enabling control of shapes, localization degree, and location of the edge states in the spectrum of the system by their power~\cite{smirnova.apr.7.021306.2020}. The impact of nonlinearity on anomalous topological states in conservative SSH arrays and formation of topological $\pi$ solitons has been studied only recently, in both theory~\cite{zhong.pra.107.L021502.2023} and experiment~\cite{arkhipova.sb.68.2017.2023}. While nonlinear $\pi$ modes are already observed in conservative systems, in non-Hermitian systems the results on such states are scarce: only experiments in linear regime are available~\cite{wu.prr.3.023211.2021}. At the same time, even in unmodulated non-Hermitian SSH arrays high-power light beams demonstrate much richer evolution scenarios~\cite{weimann.nm.16.1476.2017, xia.science.372.72.2021} in comparison with conservative SSH structures. Therefore, it is particularly interesting to study the impact of nonlinearity (including nonlinear absorption) on the dynamics of non-Hermitian anomalous topological systems. As we show below such systems allow observation of stable lasing in $\pi$ modes.

It should be mentioned that design of topological lasers is considered as one of the most promising applications of topological insulators, because such lasers are potentially resilient to disorder, they may have smaller linewidth and larger slope efficiency in comparison with their nontopological counterparts. Topological lasers have been constructed and observed on different photonic and optoelectronic platforms~\cite{schomerus.ol.38.1912.2013,jean.np.11.651.2017, bahari.science.358.636.2017, harari.science.359.eaar4003.2018, bandres.science.359.eaar4005.2018, kartashov.prl.122.083902.2019, zeng.nature.578.246.2020, zhong.lpr.14.2000001.2020, smirnova.light.9.127.2020, zhang.light.9.109.2020, kim.nc.11.5758.2020, yang.prl.125.013903.2020, yang.prx.10.011059.2020, choi.nc.12.3434.2021, zhong.apl.6.040802.2021, dikopoltsev.science.373.1514.2021, yang.np.16.279.2022, zhu.prl.129.013903.2022, ezawa.prr.4.013195.2022, wu.sa.9.eadg4322.2023, li.prl.131.023202.2023, han.light.12.1.2023}, including static SSH arrays and their two-dimensional generalizations~\cite{schomerus.ol.38.1912.2013,jean.np.11.651.2017, kim.nc.11.5758.2020, ezawa.prr.4.013195.2022, teo.prb.105.L201402.2022, wu.sa.9.eadg4322.2023, li.prl.131.023202.2023}. However, Floquet lasers have been studied so far only in system supporting unidirectional edge states~\cite{ivanov.apl.4.126101.2019}. Lasing in anomalous edge states in Floquet systems has not been reported so far, to the best of our knowledge.

Here we report on stable lasing in $\pi$ modes emerging in one-dimensional non-Hermitian SSH waveguide array. The Floquet mechanism is mimicked here by periodic out-of-phase modulations of the positions of neighboring waveguides in the array in light propagation direction. In contrast to conservative Floquet SSH systems~\cite{dallago.pra.92.023624.2015, fruchart.prb.93.115429.2016, zhang.acs.4.2250.2017, petracek.pra.101.033805.2020, cheng.prl.122.173901.2019, song.lpr.15.2000584.2021, zhong.pra.107.L021502.2023, arkhipova.sb.68.2017.2023}, we assume that gain is provided in one waveguide of the unit cell, while the other waveguide is lossy, i.e. the structure at any distance ``instantaneously" satisfies parity-time symmetry condition~\cite{weimann.nm.16.1476.2017}. Longitudinal array modulation leads to the formation of $\pi$ modes at different ends of the array, one of which is amplified, while other is damped. We show that the former mode is responsible for stable lasing when nonlinearity and nonlinear absorption are included. In this new type of Floquet topological laser spontaneous transition to lasing in $\pi$  modes is observed even for noisy inputs despite the fact that gain is present in each cell of the structure.

It should be mentioned that our system does not contain resonator as it happens in conventional lasers, where light performs multiple roundtrips in active medium. Instead, the system is formally unlimited in light propagation direction $z$ (longitudinal modulations of optical potential do not cause backward reflection because the potential is shallow). The formation of "lasing" modes in our dissipative system with linear gain and nonlinear absorption should be understood as formation of stable attractors (to which the input beam will approach after sufficiently long propagation distance, if it belongs to the basin of attractor), which, in addition, demonstrate peculiar evolution dynamics because parameters of our dissipative system change with evolution variable $z$.

\section{Results}

\subsection{Theoretical model}

\begin{figure*}[htbp]
\centering
\includegraphics[width=\textwidth]{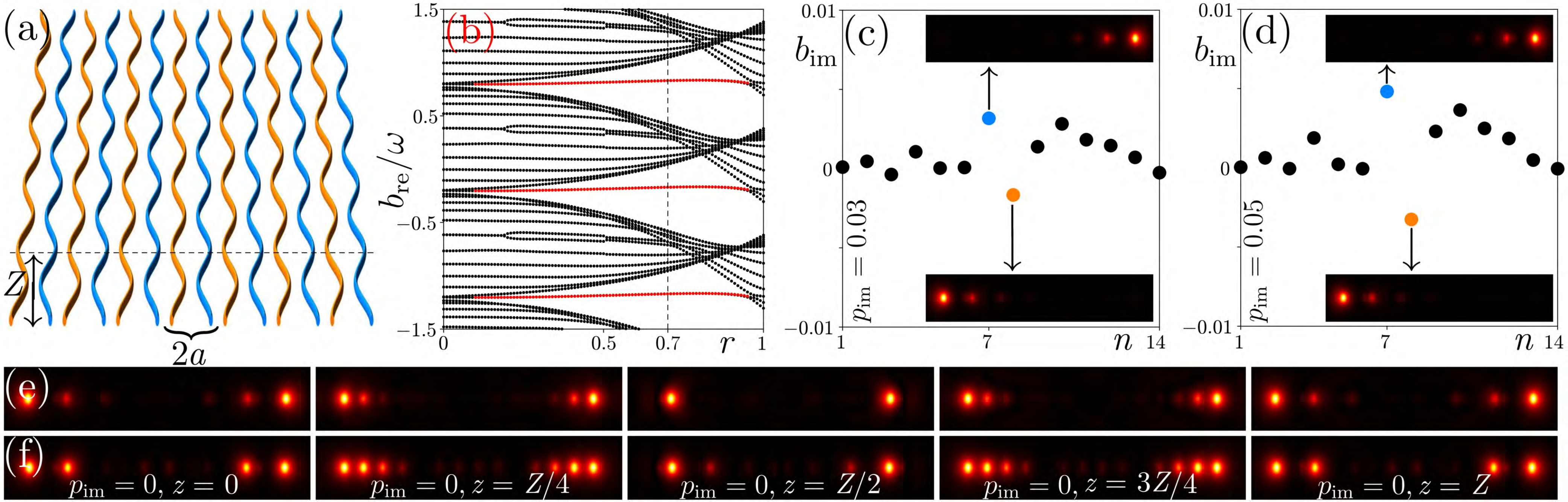}
\caption{\textbf{Non-Hermitian Floquet SSH array and its spectrum.}
(a) Schematic representation of the non-Hermitian Floquet SSH array with gain in orange waveguides and losses in blue waveguides. (b) Quasi-propagation constants of eigenmodes of the Floquet SSH array versus amplitude of waveguide oscillations at $p_\textrm{im}=0$. Three longitudinal Brillouin zones are shown. The red dots correspond to $\pi$ modes appearing in this case at both ends of the array, while the black dots correspond to bulk states. (c) Imaginary parts of quasi-propagation constants of all eigenmodes with index $n$ in the array with ${r=0.7}$ and ${p_{\rm im}=0.03}$. Field modulus distributions in $\pi$ mode that experiences attenuation (blue dot) and $\pi$ mode that is amplified (orange dot) upon propagation. They appear at different edges, as shown in the insets. (d) Same dependence as in (c), but for ${p_{\rm im}=0.05}$. 
(e) Evolution dynamics of $\pi$ modes in conservative array with ${p_{\rm im}=0}$ and ${r=0.7}$ within one longitudinal period $Z$. 
(f) Similar to (e) but for ${r=0.5}$.
$|\psi|$ distributions in (c,d,e) are shown for ${-24\le x \le 24}$ and ${-5\le y \le 5}$.}
\label{fig1}
\end{figure*}

We consider paraxial propagation of light beams in shallow waveguide arrays, whose profile changes periodically in the direction of light propagation $z$. The propagation dynamics of a light beam can be described by the normalized nonlinear Schr\"odinger-like equation
\begin{align} \label{eq1}
i \frac{\partial \psi}{\partial z} = -\frac{1}{2}\nabla^2 \psi - \mathcal{R}(x,y,z) \psi - (1+i\alpha)|\psi|^2 \psi ,
\end{align}
where ${\nabla^2=\partial_x^2+\partial_y^2}$, $x$ and $y$ are the transverse coordinates normalized to the characteristic transverse scale $r_0$, $z$ is the propagation distance normalized to the diffraction length $kr_0^2$, ${k = 2\pi n_0 /\lambda }$ is the wavenumber, $n_0$ is the unperturbed refractive index of the material,  $\lambda$ is the wavelength, and ${\psi=(k^2r_0^2n_2/n_0)^{1/2}E}$ is the normalized light field amplitude with $n_2$ being the nonlinear refractive index and $E$ being the real amplitude. The function describing the profile of $z$-modulated array ${\mathcal{R} = \mathcal{R}_{\rm re} + i \mathcal{R}_{\rm im}}$ consists of two parts, where ${\mathcal{R}_{\rm re}(x,y,z)}$ describes shallow modulation of the real part of the refractive index defining waveguides, while ${\mathcal{R}_{\rm im}(x,y,z)}$ accounts for inhomogeneous gain/losses in waveguides (i.e. modulation of the imaginary part of the refractive index). We assume that in our SSH array with two waveguides in the unit cell, one waveguide is amplifying, while other waveguide is lossy [see schematics in Fig.~\ref{fig1}(a)]. Thus, 
\begin{align}\label{eq2}
\mathcal{R}  = \sum_{n=-N}^N & \left[ (p_{\rm re}+ip_{\rm im}) e^{(- x_{n,1}^2/\sigma_x^2 - y^2/\sigma_y^2)} + \right. \notag \\
& \left. (p_{\rm re}-ip_{\rm im}) e^{( - x_{n,2}^2/\sigma_x^2 - y^2/\sigma_y^2)} \right],
\end{align}
where 
\begin{align}
x_{n,1} & = x-2na-a/2-r\cos(\omega z), \notag \\
x_{n,2} & = x-2na+a/2+r\cos(\omega z)  \notag
\end{align}
are the coordinates accounting for the fact that centers of two waveguides in the unit cell perform out-of-phase oscillations along the $x$-axis with frequency $\omega$ (period ${Z=2\pi/\omega}$) and amplitude $r$. The width of the unit cell is $2a$, where $a$ is the waveguide spacing in uniform array that one obtains at ${r=0}$, ${2N+1}$ is the number of the unit cells in the array, 
\[
p_{\rm re}+ip_\textrm{im} = \frac{k^2 r_0^2}{n_0} (\delta n_\textrm{re}+i\delta n_\textrm{im}),
\] 
where $\delta n_\textrm{re}+i\delta n_\textrm{im}$ is the amplitude of complex refractive index modulation (i.e., we assume that both real and imaginary parts of linear electric susceptibility of the material are spatially modulated defining shallow complex modulation  $\delta n_\textrm{re}+i\delta n_\textrm{im}$ of the refractive index, $\sigma_{x,y}$ determines the width of the waveguides, and $\alpha$ is the coefficient of nonlinear absorption arising from all sources (including gain saturation). Notice that the conditions
\begin{align}
& \mathcal{R}_\textrm{re}(x,y,z)=\mathcal{R}_\textrm{re}(-x,y,z),\notag \\
& \mathcal{R}_\textrm{im}(x,y,z)=-\mathcal{R}_\textrm{im}(-x,y,z)
\end{align}
are satisfied at any distance $z$, i.e. the optical potential is ``instantaneously" $\mathcal{PT}$-symmetric along the $x$ axis, as one can see from the expression (\ref{eq2}) for complex function $\mathcal{R}(x,y,z)$, where coordinates $x_{n,1}, x_{n,2}$ are $z$-dependent. In schematic illustration of the array in Fig.~\ref{fig1}(a), the orange and blue colors represent amplifying and lossy waveguides, respectively. The inter- and intra-cell spacings in this system change periodically, switching it from instantaneous topological (inter-cell coupling is stronger than intra-cell one) into instantaneous non-topological (inter-cell coupling is weaker than intra-cell one) phases, i.e. in each phase the arrays spends exactly half of the $Z$ period.

The normalization of the light field amplitude used in Eq.~(\ref{eq1}) is such that the coefficient in front of conservative cubic nonlinear term equals to 1. In this case, the dimensionless beam intensity $|\psi|^2$ corresponds to real intensity
\[
I = |E|^2 = \frac{n_0}{k^2 r_0^2 n_2} |\psi|^2  .
\]
For example, for waveguide arrays fabricated in the AlGaAs material the unperturbed refractive index ${n_0\approx3.39}$, 
while typical value of nonlinear refractive index ${n_2 \approx 1.5 \times 10^{-13}\, \rm cm^2/W}$~\cite{aitchison.jqe.33.341.1997, gehrsitz.jap.87.7825.2000, mobini.micro.13.991.2022}. 
In this case, for the characteristic transverse scale ${r_0=10\,\mu \rm m}$ and ${\lambda=1550\,\rm nm}$ the dimensionless intensity ${|\psi|^2\sim 1}$ corresponds to real peak intensity ${\sim1.2\times 10^9\,\rm  W/cm^2}$. 
The relation between total beam power and the dimensionless power $U=\int\hspace{-0.5em}\int{|\psi|^2dxdy}$ is then given by
\[
\int\hspace{-0.5em}\int I dX dY = \frac{n_0}{k^2 n_2} \int\hspace{-0.5em}\int |\psi|^2 dxdy = \frac{n_0}{k^2 n_2} U,
\]
where ${(X,Y)=(xr_0,yr_0)}$ are the dimensional transverse coordinates. 
Further we assume that ${p_{\rm re}=4.5}$ (that corresponds to the refractive index modulation depth ${\delta n_\textrm{re} \sim 8.08\times 10^{-4}}$), ${p_{\rm im} \sim 0.05}$ that corresponds to linear gain/loss coefficient ${2p_{\rm im}/kr_0^2\sim 0.728\,\rm cm^{-1}}$, 
waveguide spacing ${a=3}$ ($30\, \mu \rm m$), waveguide widths ${\sigma_x=0.25}$ ($2.5\,\mu \rm m$), ${\sigma_y=0.75}$ (${7.5\,\mu \rm m}$), 
and longitudinal period ${Z=29}$ ($40\,\rm mm$). The value of $\alpha$ is connected with the two-photon absorption coefficient $\beta_2$ as ${\alpha=\beta_2\lambda/4\pi n_2}$~\cite{yang.ol.17.710.1992}, 
so ${\alpha=0.3}$ corresponds to ${\beta_2 \sim 3.6 \times 10^{-9}\, \rm cm/W}$ (that is equivalent to "imaginary part" of the nonlinear refractive index ${\alpha n_2\sim 4.5 \times 10^{-14}\, \rm cm^2/W}$). 
Technically the realisation of inhomogeneous gain/losses may be achieved by inhomogeneous doping of the material with active ions or utilization of inhomogeneous pump. 
We would like to note that we deal with normalized quantities to ensure that the model Eq.~(\ref{eq1}) 
is general and can be potentially applied to different materials and settings.

\subsection{ $\pi$ modes in non-Hermitian Floquet SSH waveguide array}

Exact Floquet states of the \textit{nonlinear} system governed by Eq. (\ref{eq1}) (i.e. states \textit{exactly} reproducing their profiles after each longitudinal period $Z$) can be written in the form 
\[\psi=u(x, y, z) e^{i b z}, \]
where $b$ is the real-valued quasi-propagation constant and ${u(x, y, z)=u(x, y, z+Z)}$ is the complex $Z$-periodic function that solves the equation
\begin{equation}\label{eq4}
b u=\frac{1}{2}\nabla^2 u + \mathcal{R} u+i \frac{\partial u}{\partial z}+(1+i\alpha)|u|^2 u.
\end{equation}
Before discussing its possible solutions, it is instructive to consider linear spectrum of this system, neglecting last nonlinear term in Eq.~(\ref{eq4}). In a general case with complex $\mathcal{R}$ the quasi-propagation constants of linear eigenmodes also become complex: ${b=b_{\rm re}+ i b_{\rm im}}$. In conservative system with ${\mathcal{R}_\textrm{im}=0}$ the quasi-propagation constants are real. Due to $Z$ periodicity of the function $\mathcal{R}$, the quasi-propagation constants are defined in the first longitudinal Brillouin zone ${b\in[-\omega/2,\omega/2]}$ and the spectrum is periodically replicated in $b$ with period $\omega$. Linear spectrum of such conservative Floquet SSH array can be found using propagation and projection method~\cite{smirnova.apr.7.021306.2020, ivanov.apl.4.126101.2019, ren.chaos.166.113010.2023, arkhipova.sb.68.2017.2023} and it is presented in Fig.~\ref{fig1}(b), where red dots correspond to the $\pi$ modes simultaneously appearing at both edges of the array due to its symmetry, and black dots correspond to the bulk modes. Notice that $\pi$ modes emerge around the points of overlap of the Floquet replicas of the same band, in the gap that opens due to waveguide oscillations, whose width initially increases with increase of the amplitude of waveguide oscillations $r$. The example of evolution of the $\pi$ mode in conservative system at ${r=0.7}$ is shown in Fig.~\ref{fig1}(e) for five selected distances within one $Z$ period. Such states strongly oscillate, but nevertheless they \textit{exactly} reproduce their profile after each period, when Floquet system is conservative.
The $\pi$ mode at ${r=0.5}$ is shown in Fig.~\ref{fig1}(f). Clearly, the localization of the $\pi$ mode in Fig.~\ref{fig1}(e) is better than in Fig.~\ref{fig1}(f), so the amplitude of waveguide oscillations $r$ notably affects the localization of the $\pi$ modes~\cite{arkhipova.sb.68.2017.2023}.
Indeed, a general property of all in-gap states is that their localization increases close to the center of the gap and decreases close to its edges. Moreover, wider gap always provides better localization for in-gap modes. Because $\pi$ modes remain close to the center of the gap for a broad range of oscillation amplitudes $r$ and because gap width initially also increases with $r$, as one can see from Fig. \ref{fig1}(b), we observe the progressive enhancement of localization of $\pi$ modes at least up to $r=0.8$.

For nonzero $p_{\rm im}$ Floquet SSH array becomes non-Hermitian. Corresponding quasi-propagation constants $b$ of linear eigenmodes become complex. For sufficiently small $p_\textrm{im}$, ${0<p_{\rm im} \ll 1}$, the propagation and projection method can still be used to calculate quasi-propagation constants $b$ and it produces the same dependence $b_{\rm re}(r)$ for real part of the quasi-propagation constant as the one shown in Fig. \ref{fig1}(b). However, even small imaginary part $\sim p_{\rm im}$ of the optical potential has dramatic impact on the profiles of $\pi$ modes: they now localize at the opposite ends of the array, see the insets in Figs.~\ref{fig1}(c) and \ref{fig1}(d). Nonzero imaginary part of the quasi-propagation constant for a given mode can be also extracted using its evolution on one $Z$-period:
\begin{equation}\label{eq3}
b_{\rm im} = - \frac{1}{Z} \ln \frac{|\psi_{\rm out}|}{|\psi_{\rm in}|},
\end{equation}
where $\psi_{\rm in}$ and $\psi_{\rm out}$ are the input and output field distributions of the mode (this procedure is valid for small $p_{\rm im}$ as long as hybridization of modes does not occur). In Figs.~\ref{fig1}(c) and \ref{fig1}(d), we show the imaginary parts of quasi-propagation constants of all modes of the array with 7 cells at ${r=0.7}$ for ${p_{\rm im}=0.03}$ and ${p_{\rm im}=0.05}$, respectively. The blue and orange dots correspond to $\pi$ modes localized at the opposite edges of the array: the mode corresponding to the orange dot is amplified in the course of propagation, while the mode corresponding to the blue dot is damped. As one can see, $\pi$ modes experience faster growth/attenuation in comparison with bulk states. This means that after sufficiently long evolution, the amplified $\pi$ mode will dominate over all other modes of the system, even if it emerges spontaneously from noise. Similar results are obtained for any number of cells in the array. It should be also mentioned that passive setting with different losses in two waveguides in the unit cell presumably may also support non-degenerate (although, of course, purely lossy) $\pi$ modes at the opposite edges of the array.

\begin{figure*}[htbp]
	\centering
	\includegraphics[width=\textwidth]{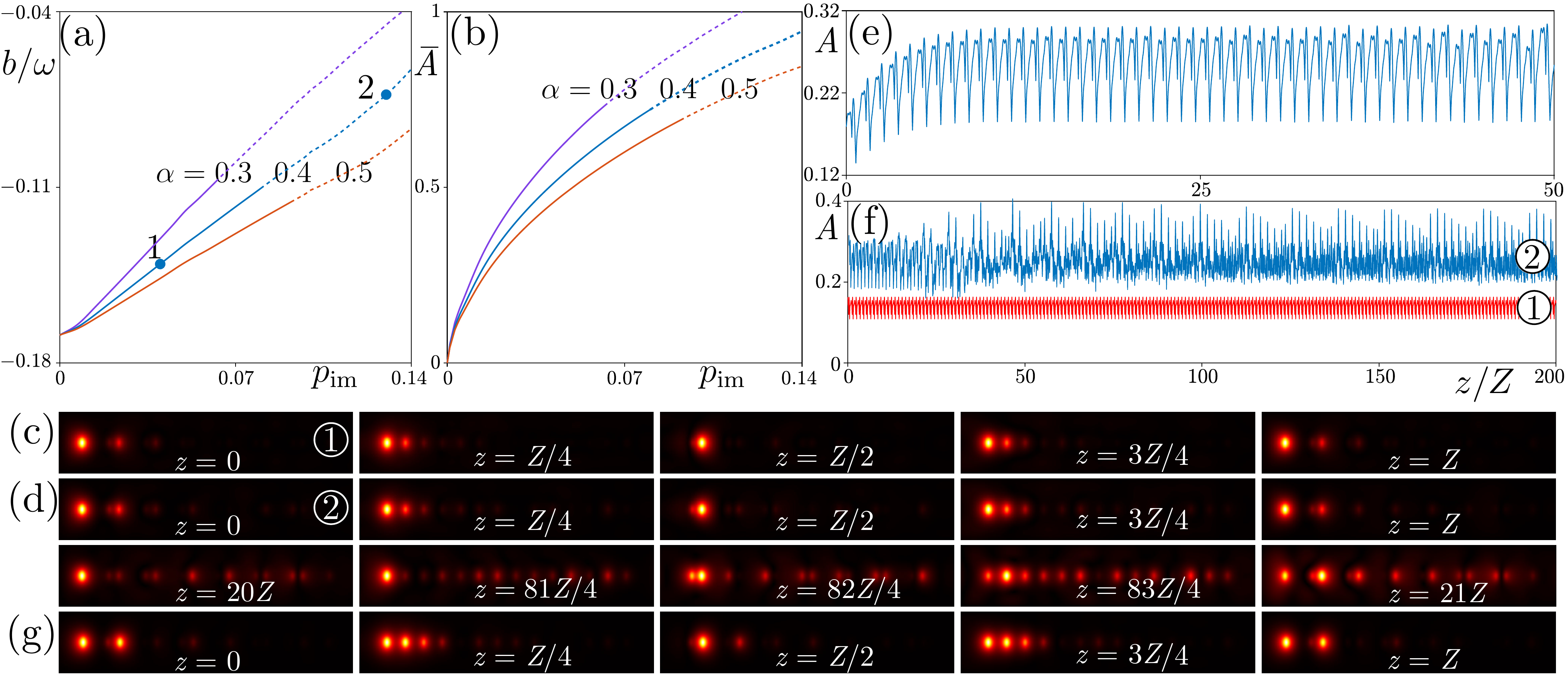}
	\caption{\textbf{$\pi$ mode lasing.} Quasi-propagation constant (a) and averaged peak amplitude $\bar A$ (b) of the nonlinear $\pi$ mode versus ${p_{\rm im}}$ for ${r=0.7}$ and different values of $\alpha$. (c) Field modulus distributions at different distances within one longitudinal period showing stable evolution of nonlinear $\pi$ mode numbered 1 in (a) for ${\alpha=0.4}$ and ${p_{\rm im}=0.04}$. (d) Panels in the first row are same as in (c) but for the unstable $\pi$ mode numbered 2 in (a) for $\alpha=0.4$ and ${p_{\rm im}=0.13}$. Panels in the second row show its unstable profiles. (e) Peak amplitude versus $z$ dependence showing establishment of lasing regime at the initial stages of evolution for ${p_{\rm im}=0.13}$ that is later followed by the formation of breather. (f) Peak amplitude versus distance for stable nonlinear $\pi$ mode at ${p_{\rm im}=0.04}$ (red curve) and unstable $\pi$ mode at ${p_{\rm im}=0.13}$  (blue curve). 
	(g) Same as (c) but for ${r=0.5}$. Distributions in (c,d,g) are shown within window ${-24\le x \le 24}$ and ${-5\le y \le 5}$.
	}
	\label{fig2}
\end{figure*}

\subsection{$\pi$ mode lasing}

To prevent unlimited growth of amplitude of the $\pi$ mode that experiences amplification and to achieve steady lasing regime, where this state shows exactly periodic oscillations, remaining localized near one edge of the structure, we now take into account focusing nonlinearity and nonlinear absorption in Eq.~(\ref{eq1}). To obtain steadily evolving \textit{nonlinear} $\pi$ modes we start with linear growing $\pi$ mode and propagate it in the frames of Eq.~(\ref{eq1}) up to a very large distance, waiting for the establishment of steady state with exactly periodic behavior of peak amplitude ${A(z)=\max|\psi|}$ of emerging nonlinear state. Since the amplitude $A$ oscillates during propagation, it is convenient to introduce its period-averaged value 
\[
\bar{A}=\frac{1}{Z}\int_z^{z+Z} A(z) dz.
\]
Real-valued quasi-propagation constant of solution in steady-state regime can be found as 
\[
b={\rm Im} \left\{ \frac{1}{Z} \ln \frac{\psi(z+Z)}{\psi(z)} \right\}. 
\]
By gradually increasing the amplitude of gain/losses $p_{\rm im}$ we obtained the families $b(p_\textrm{im})$, $\bar{A}(p_\textrm{im})$ of lasing $\pi$ modes for different values of the absorption coefficient ${\alpha=0.3}$, $0.4$, and $0.5$ that are depicted in Figs.~\ref{fig2}(a) and \ref{fig2}(b). As one can see from these figures, lasing in this model is formally thresholdless, i.e. it occurs for any ${p_\textrm{im}>0}$. In reality, however, since all active materials possess certain background losses at low pump levels, the pump would need first to compensate for such background losses, before reaching the parameter region (gain level) at which lasing in $\pi$ modes would occur. When ${p_\textrm{im} \to 0}$ the amplitude of lasing mode vanishes, while its quasi-propagation constant coincides with quasi-propagation constant of linear $\pi$ mode in conservative array. In this sense, lasing modes bifurcate from modes of conservative array and, importantly, their quasi-propagation constants remain inside topological gap of conservative system that guarantees the absence of coupling with bulk modes. Both $b$ and $\bar{A}$ increase away from bifurcation point. For a given $p_{\rm im}$, increasing nonlinear absorption coefficient $\alpha$ leads to decrease of both $b$ and $\bar{A}$.

We found that stable lasing occurs in certain interval of $p_\textrm{im}$ values adjacent to the bifurcation point, while for larger $p_\textrm{im}$ values lasing becomes unstable/multimode and may be accompanied by the formation of breathers, whose periodicity substantially exceeds the period $Z$ of the array oscillations. The regions, where stable lasing occurs are marked in Figs.~\ref{fig2}(a) and \ref{fig2}(b) with solid lines, while regions where lasing becomes unstable are shown with dashed lines. The dynamics of establishment of stable and unstable regimes are illustrated in Fig.~\ref{fig2}(f) with red and blue curves, respectively. While in stable lasing regime the evolution of amplitude $A(z)$ is exactly periodic at large distances with period equal to $Z$ (red curve), in the unstable regime [see blue curve and its zoom in Fig.~\ref{fig2}(e)] the mode first reaches the stage, where $A$ oscillates nearly periodically, but then slow transition into breathing regime occurs, with period nearly 10 times larger than waveguide oscillation period $Z$ [see Fig.~\ref{fig2}(f)]. The dynamics of stable and unstable lasing $\pi$ modes corresponding to the dots 1 and 2 in Fig.~\ref{fig2}(a), is illustrated in Figs.~\ref{fig2}(c) and \ref{fig2}(d), respectively, on the distance corresponding to one array period $Z$. The recovery of the initial field distribution in stable nonlinear state at ${z=Z}$ is exact despite strong shape transformations during one period [Fig.~\ref{fig2}(c)]. In the unstable regime [Fig.~\ref{fig2}(d)] no exact recovery of initial distribution after one period occurs, but because breathing period in this case substantially exceeds $Z$, the differences in shapes at $z$ and ${z+Z}$ are rather small (mainly peak amplitude of the mode changes). On this reason, the first row in Fig.~\ref{fig2}(d) illustrates evolution of mode corresponding to dot 2 only on the first longitudinal array period $Z$ to stress that it nearly recovers its shape. However, in second row of Fig.~\ref{fig2}(d) we show field modulus distributions in this state after many longitudinal periods, from which one can observe that this solution is indeed unstable. Notice that because in the unstable regime the values of $b$ and $\bar{A}$  usually slightly change from period to period, we presented in Figs.~\ref{fig2}(a) and \ref{fig2}(b) corresponding values averaged over $100$ periods $Z$. Notice that for even larger ${p_\textrm{im}>0.15}$ values in comparison with the regime, where regular breathing occurs, one observes transition to irregular amplitude oscillations. It should be also mentioned that the domain of $p_\textrm{im}$ values, where stable lasing occurs gradually expands with increase of the nonlinear absorption coefficient $\alpha$. In Fig.~\ref{fig2}(g), we also show an example of lasing $\pi$ mode at ${r=0.5}$, for the same set of other parameters as in Fig.~\ref{fig2}(c). By comparing Figs.~\ref{fig2}(c) and \ref{fig2}(g), one finds that the localization of the lasing $\pi$ mode can be flexibly controlled by the amplitude of waveguide oscillations $r$.

\begin{figure*}[htbp]
	\centering
	\includegraphics[width=1.63\columnwidth]{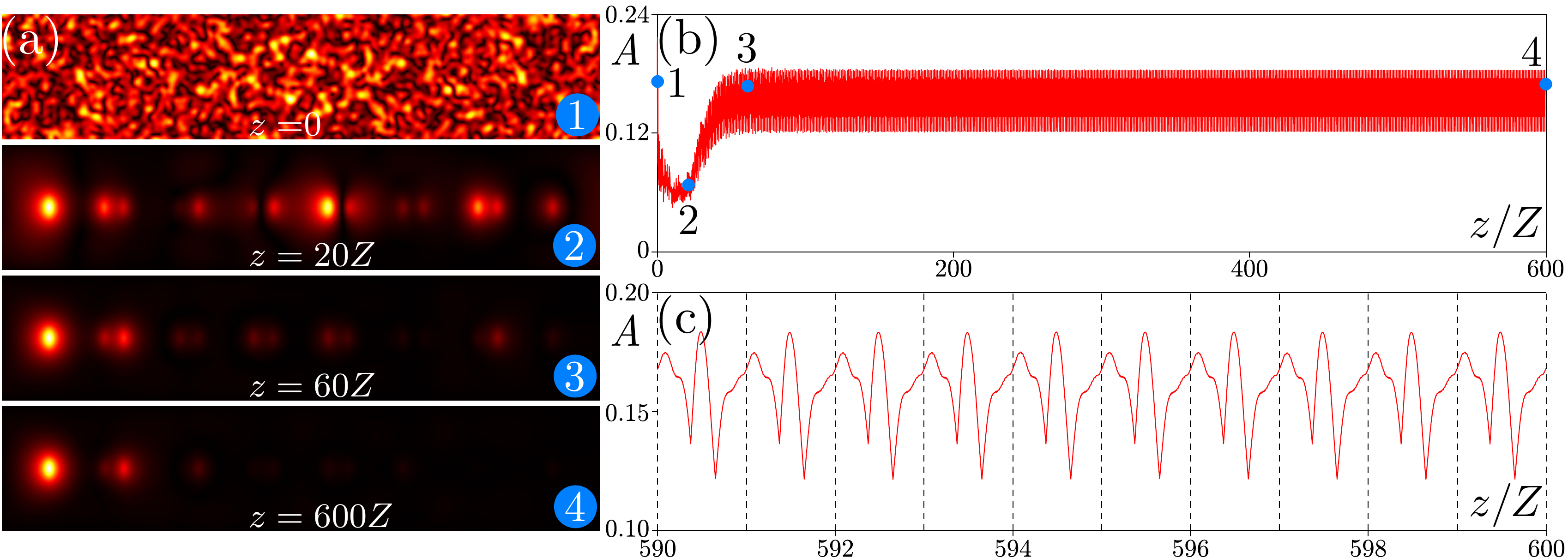}
	\caption{\textbf{Excitation of the lasing $\pi$ mode from a random noise.} (a) Field modulus distributions at selected distances indicated by the numbers in (b). (b) Peak amplitude of the field versus propagation distance with four dots at ${z=0}$ (dot 1), ${z=20Z}$ (dot 2), ${z=60Z}$ (dot 3), and ${z=600Z}$ (dot 4). (c) Zoom of the dependence in (b) stressing exactly periodic variation of amplitude on 10 longitudinal periods in steady-state regime (dashed lines are guides for the eye, indicating each period). In all cases ${p_{\rm im}=0.05}$ and ${\alpha=0.4}$. $|\psi|$ distributions in (a) are shown within ${-24\le x \le 24}$ and ${-5\le y \le 5}$ window.
	}
	\label{fig3}
\end{figure*}

\subsection{Excitation of $\pi$ modes from noisy inputs}

As mentioned above, one of linear $\pi$ modes of dynamical complex landscape $\mathcal{R}$ features fastest amplification among all modes of the system. Therefore, in proper parameter range, for excitations with broadband input noise, corresponding $\pi$ mode is likely to win the competition with all other modes, resulting in single-mode lasing in the nonlinear regime. This fact is demonstrated in Fig.~\ref{fig3}, where we show such excitation dynamics and transition to stable lasing. One can see that after initial transient stage, where input noise (see distribution corresponding to dot 1), excites the combination of several modes including edge and bulk ones (dot 2), the edge mode starts dominating and remains the only excited state at sufficiently large distances (dots 3 and 4). Single-mode lasing is confirmed by exactly periodic dependence of peak amplitude $A$ on distance $z$ already after ${z>50Z}$, see Fig.~\ref{fig3}(b) and its zoom in Fig.~\ref{fig3}(c). Therefore, nonlinear $\pi$ modes represent stable attractors with sufficiently large basin in this dissipative system. We have checked that similar picture holds for other values of $p_{\rm im}$ and $\alpha$ parameters, where the system supports stable single-mode lasing, see Figs.~\ref{fig2}(a) and \ref{fig2}(b).
We would like to mention that since in our dissipative system $\pi$ modes represent stable attractors, whose parameters are determined exclusively by the gain/loss amplitude $p_{\rm im}$ and nonlinear absorption coefficient $\alpha$, they are excited from a very broad range of input conditions, should it be noise with different amplitudes or localized input beams, but for fixed $p_{\rm im}$ and $\alpha$ the outcome of the excitation is typically the same – stably oscillating $\pi$-mode – as long as the input remains within the basin of attractor.

In this article we only consider spatial dynamics of the system assuming monochromatic regime and omitting dispersion effects and temporal dynamics of the gain medium (like evolution of populations of levels providing gain, etc). Inclusion of temporal dynamics of course may bring new effects, and it can even lead to potential instabilities~\cite{longhi.epl.122.14004.2018, longhi.ol.44.287.2019}. While in topological lasers based on photonic crystals linewidth narrowing in frequency domain is manifested as a sharp concentration of the spectrum of radiation around eigenfrequency $\omega$ of in-gap topological mode, in which lasing occurs, in our case lasing is also manifested in the formation of nonlinear state with single well-defined quasi-propagation constant $b$ in topological gap. When such lasing is single-mode, only one such state is excited, but in multimode lasing regime at sufficiently high gain amplitudes the excitation of dynamical oscillating states with contributions from the limited interval of quasi-propagation constants $b$ may be possible (that can be considered as an analogue of the broadening of the linewidth).

\subsection{Discussions}

From the point of view of potential experimental realisation of the setting considered here, we would like to mention that in addition to AlGaAs material mentioned above, the waveguide arrays exhibiting gain and losses can be fabricated in numerous ways. They include the technique of direct fs-laser writing of waveguides that is demonstrated to be efficient in a broad range of transparent materials~\cite{gattass.np.2.219.2008}, including those containing amplifying dopants. To the best of our knowledge, various waveguides were already realized in Er-doped active phosphate~\cite{chiodo.ol.11.1651.2006}, silicate~\cite{thomson.apl.87.121102.2005}, tellurite~\cite{fernande.oe.16.15198.2008}, and Baccarat glasses~\cite{vishnubhatla.jpdap.42.205106.2009}, and also in lithium niobate~\cite{burghoff.apl.89.081108.2006} allowing realization of inhomogeneous parametric gain used for observation of parity-time symmetry~\cite{ruter.np.6.192.2010}. 

\section{Conclusions}

Summarizing, we have reported on stable lasing in $\pi$ modes in non-Hermitian Floquet SSH waveguide arrays. Despite the fact that gain is provided in each unit cell in such structures, $\pi$ modes can be stable attractors in proper range of gain/losses amplitudes, and win the competition with other modes, leading to stable single-mode lasing regime. Corresponding lasing modes have quasi-propagation constants in topological gap, indicating on their topological protection, resilience to noise and disorder. Spatial localization degree of lasing modes can be controlled in this system by changing amplitude of transverse waveguide oscillations. Our results open new avenue for utilization and control of topological $\pi$ modes in non-Hermitian Floquet systems, and may be interesting for design of compact laser sources.

\begin{acknowledgments}		
This work was supported by the Natural Science Basic Research Program of Shaanxi Province (2024JC-JCQN-06), the National Natural Science Foundation of China (12074308), partially by the research project FFUU-2024-0003 of the Institute of Spectroscopy of the Russian Academy of Sciences, the Russian Science Foundation (grant 24-12-00167), and the Fundamental Research Funds for the Central Universities (xzy022023059).
\end{acknowledgments}

\section*{DATA AVAILABILITY}
The data that support the findings of this study are available
from the corresponding author upon reasonable request.

\section*{REFERENCES}
%

\end{document}